\documentstyle[aps,prl]{revtex}

\newcommand{\BE}{\begin{equation}}
\newcommand{\EE}{\end{equation}}
\newcommand{\BA}{\begin{eqnarray}}
\newcommand{\EA}{\end{eqnarray}}

\begin{document}
\draft

\twocolumn[\hsize\textwidth\columnwidth\hsize\csname@twocolumnfalse\endcsname
\title{Acoustic Attenuation by Two-dimensional Arrays of Rigid Cylinders}
\author{You-Yu Chen and Zhen Ye}
\address{Wave Phenomena Laboratory, Department of Physics,
National Central University, Chungli, Taiwan 32054}
\date{January 30, 2001}

\maketitle

\begin{abstract}

In this Letter, we present a theoretical analysis of the acoustic
transmission through two-dimensional arrays of straight rigid
cylinders placed parallelly in the air. Both periodic and
completely random arrangements of the cylinders are considered.
The results for the sound attenuation through the periodic arrays
are shown to be in a remarkable agreement with the reported
experimental data. As the arrangement of the cylinders is
randomized, the transmission is significantly reduced for a wider
range of frequencies. For the periodic arrays, the acoustic band
structures are computed by the plane-wave expansion method and are
also shown to agree with previous results.

\end{abstract}

]

When propagating through media containing many scatterers, waves
will be scattered by each scatterer. The scattered waves will be
scattered again by other scatterers. This process is repeated to
establish an infinite recursive pattern of rescattering between
scatterers, forming a multiple scattering process
\cite{Ishimaru,Ye0}. Multiple scattering of waves is responsible
for a wide range of fascinating phenomena, including twinkling
light in the evening sky, modulation of ocean ambient
sound\cite{Ye1}, acoustic scintillation from fish
schools\cite{YT}. On smaller scales, phenomena such as white
paint, random laser\cite{Laser}, electron transport in impured
solids\cite{Im} are also results of multiple scattering. When
waves propagate through media with periodic structures, the
multiple scattering leads to the ubiquitous phenomenon of band
structures. That is, waves can propagate in certain frequency
ranges and follow a dispersion relation, while within other
frequency regimes wave propagation is stopped. The former ranges
are called allowed bands and the latter the forbidden bands. In
certain situations, the inhibition of wave propagation occurs for
all directions, leading to the phenomenon of complete band gaps.

The wave dispersion bands are first studied for electronic waves
in solids, providing the basis for understanding the properties of
conductors, semi-conductors, and insulators\cite{Kittel}. In late
1980s, it became known that such a wave band phenomenon is also
possible for electro-magnetic waves in media with periodically
modulated refractive-indices\cite{Optic}. Since then, optical wave
bands have been extensively studied, yielding a rich body of
literature. The theoretical calculations have proven to match well
with the experimental observations\cite{Exp1}.

In contrast, research on acoustic wave band structures has just
started. Although theoretical computations of band structures have
been documented for periodic acoustic structures\cite{Kush}, the
experimental work was only recent, and to date only a limited
number of measurements has been reported. One of the first
observations was made on acoustic attenuation by a
sculpture\cite{Sculpture}. The authors obtained a sound
attenuation spectrum, which was later verified by the band
structure computation\cite{Kush2}. Recently, acoustic band
structures have been further measured for acoustic transmission
through two-dimensional (2D) periodic arrays of rigid cylinders
placed in the air \cite{Sanchez}. The authors demonstrated the
properties of sound attenuation along two high-symmetry directions
of the Brillouin zone of the arrays. They also observed a peculiar
effect of deaf bands; within the bands, in spite of non-zero band
states, wave propagation is prohibited due to particular symmetry
of the states\cite{Sanchez}.

The main purpose of this Letter is to provide a theoretical
investigation of sound transmission by 2D arrays of rigid
cylinders in air in line with the experiment of \cite{Sanchez},
thus providing a direct comparison of attenuation between theory
and experiment. Such a direct comparison is relatively scarce in
the literature. We note that the comparison between the
attenuation spectrum and the dispersion bands is of indirect
nature, as the two are not necessarily one to one correspondent,
as seen, for instance, when some seemingly allowed bands are
actually deaf to wave transmission\cite{Sanchez}. This will be
further clarified in the later results. Another goal is to study
how the randomness affects the acoustic transmission, so to gain
some insight into the connection between the forbidden bands and
wave localization in 2D\cite{Ye2}. For the purposes, we adopt a
self-consistent multiple scattering theory\cite{Twersky}.

Consider $N$ straight cylinders located  at $\vec{r}_i$ with $i=1,
2, \cdots, N$ to form either a regular lattice or a random array
perpendicular to the $x-y$ plane; the regular arrangement can be
adjusted to comply with the experiment\cite{Sanchez}. The
cylinders are along the $z$-axis. An acoustic source transmitting
monochromatic waves is placed at $\vec{r}_s$, some distance from
the array. The scattered wave from each cylinder is a response to
the total incident wave composed of the direct wave from the
source and the multiply scattered waves from other cylinders. The
final wave reaches a receiver located at $\vec{r}_r$ is the sum of
the direct wave from the source and the scattered waves from all
the cylinders. Such a scattering problem can be formulated {\it
exactly} in the cylindrical coordinates, following
Twersky\cite{Twersky}. While the details are in \cite{Yep}, the
essential procedure is presented below.

The scattered wave from the $j$-th cylinder ($j = 1,2,\cdots, N$)
can be written as \BE \label{eqps1} p_s(\vec{r}, \vec{r}_j) =
\sum_{n=-\infty}^{\infty} \mbox{i}\pi A_n^j H_n^{(1)}(k|\vec{r} -
\vec{r}_j|)e^{\mbox{i}n\phi_{\vec{r}- \vec{r}_j}}, \EE where
$\mbox{i} = \sqrt{-1}$, $H_n^{(1)}$ is the $n$-th order Hankel
function of the first kind, $\phi_{\vec{r}- \vec{r}_j}$ is the
azimuthal angle of the vector $\vec{r}- \vec{r}_i$ relative to the
positive $x$-axis. The total incident wave around the $i$-th
cylinder ($i=1,2,\cdots, N; i\neq j$) is \BE \label{eqpin1}
p_{in}^i(\vec{r}) = p_0(\vec{r}) + \sum_{j=1,j\neq i}^N
p_s(\vec{r}, \vec{r}_j), \EE which can be expressed again in terms
of a modal series \BE \label{eqpin2} p_{in}^i(\vec{r}) = \sum_{n =
-\infty}^\infty B_n^i J_n(k|\vec{r} -
\vec{r_i}|)e^{\mbox{i}n\phi_{\vec{r} - \vec{r_i}}}. \EE The
expansion is in terms of Bessel functions of the first kind $J_n$
to ensure that $p_{in}^i(\vec{r})$ does not blow up as $\vec{r}
\rightarrow \vec{r_i}$.

To solve for $A_{n}^i$ and $B_n^i$, we express the scattered wave
$p_s(\vec{r}, \vec{r_j})$, for each $j \neq i$, in terms of the
modes with respect to the $i$-th scatterer by the addition theorem
for the Bessel functions\cite{addition}. The resulting formula for
the scattered wave $p_s(\vec{r}, \vec{r_j})$ is \BE \label{eqps2}
p_s(\vec{r}, \vec{r_j}) = \sum_{n=-\infty}^\infty C_n^{j, i}
J_n(k|\vec{r} - \vec{r_i}|)e^{\mbox{i}\phi_{\vec{r} - \vec{r_i}}},
\EE with \BE C_n^{j,i} = \sum_{l=-\infty}^\infty \mbox{i}\pi A_l^j
H_{l-n}^{(1)}(k|\vec{r_i} - \vec{r_j}|)
e^{\mbox{i}(l-n)\phi_{\vec{r_i} - \vec{r_j}}}. \label{Cn}\EE The
direct incident wave around the location of the $i$-th cylinder
can be expressed in the Bessel function expansion with respect to
the coordinates centered at $\vec{r}_i$ \BE \label{eqp0exp}
p_0(\vec{r}) = \sum_{l=-\infty}^{\infty} S_l^i J_l(k|\vec{r} -
\vec{r_i}|) e^{\mbox{i}l\phi_{\vec{r} - \vec{r_i}}}, \EE with the
known coefficients $$ S_l^i = \mbox{i}\pi H_{-l}^{(1)}(k|\vec{r_i}
-\vec{r}_s|) e^{-\mbox{i}l\phi_{\vec{r_i}}}.$$

Matching the coefficients in equation (\ref{eqpin1}), using
equations (\ref{eqpin2}), (\ref{eqps2}) and (\ref{eqp0exp}), we
have \BE\label{eqmatrix1} B_n^i = S_n^i + \sum_{j=1,j\neq i}^N
C_n^{j, i}.\EE At this stage, both the $S_n^i$ are known, but both
$B_n^i$ and $A_l^j$ are unknown.  Boundary conditions will give
another equation relating them. The boundary conditions state that
the pressure and the normal velocity be continuous across the
interface between a scatterer and the surrounding medium. After a
deduction, we obtain \BE \label{BA} B_n^i = \mbox{i}\pi\Gamma_n^i
A_n^i, \EE where \BE \Gamma_n^i = \frac{H_n^{(1)}(k a^i) J_n'(k
a^i/h^i) - g^i h^i H_n^{(1)\prime}(k a^i) J_n(k a^i/h^i)} {g^i h^i
J_n'(k a^i) J_n(k a^i/h^i) - J_n(k a^i) J_n'(k a^i/h^i)}. \EE Here
the primes refer to taking derivative, $a^i$ is the radius of the
$i$-th cylinder, $g^i = \rho_1^i/\rho$ is the density ratio, and
$h^i = k/k_1^i = c_1^i/c$ is the sound speed ratio for the $i$-th
cylinder.

The unknown coefficients $A_n^{i}$ and $B_n^j$ can be inverted
from Eqs.~(\ref{Cn}), (\ref{eqmatrix1}), and (\ref{BA}). Once
$A_n^{i}$ are determined, the transmitted wave at any spatial
point is given by \BE \label{final} p(\vec{r}) = p_0(\vec{r}) +
\sum_{i=1}^N \sum_{n=-\infty}^{\infty} \mbox{i}\pi A_n^i
H_n^{(1)}(k|\vec{r} - \vec{r}_i|)e^{\mbox{i}n\phi_{\vec{r}-
\vec{r}_i}}. \EE The acoustic intensity is represented by the
squared module of the transmitted wave. When the cylinders are
placed regularly, we can also obtain the band structures by the
plane wave method well documented in \cite{Kush}. The programs
used are identical to that used for computing the acoustic bands
in the regular arrays of air-cylinders in water\cite{Ye3}

Numerical computation has been carried out for the experimental
situations\cite{Sanchez} and also for an random array of the
cylinders. In the simulation, all the cylinders are assumed to be
the same, in accordance with the experiment. Moreover, the radii
of the cylinders and the lattice constants are also taken from the
experiment. Several values for the acoustic contrasts between the
cylinder and the air were used in the initial stage of
computation. We found that the results are in fact insensitive to
this factor as long as the contrasts exceed a certain value. This
agrees with the experimental observation. In the results shown
below, we use $g = h = 20$ as the values for the acoustic
contrasts. In the computation, we allow the number of the total
cylinders to vary from 100 to 500, in line with the experiment. In
the particular results shown later, we assume that the cylinders
are placed within a rectangular area of 8 $\times$ 40 of lattice
domain. The source and receiver are placed about one lattice
constant away from the long side of the array so to minimize the
effect due to the finite sample size. As we do not know the
specifications for the transmitter and the receiver, we assume an
omni-directional transmitter as the acoustic source located on one
side of the array of the cylinders, whereas an omni-directional
receiver is placed on the other side to receive the propagated
waves.

In Fig.~1, we show the relative attenuation ($\sim -\ln |p|^2$)
spectra for various square lattices for acoustic transmission
along the [100] direction. The parameters for the four cases
considered are adopted from the experiment\cite{Sanchez}. We
observe a robust attenuation peak located around 1.5 kHz for all
the situation. By eye inspection of Fig.~1 and Fig~1 in
\cite{Sanchez}, the agreement between the theoretical and the
measured results are good, particularly in view of the finite
dimension of the arrays and no adjustable parameters. The height
of the attenuation peaks depend on the locations of either the
receiver and the transmitter, on the outer boundary of the
cylinder arrays, and on the number of cylinders. Nevertheless, the
theoretical results describe quantitatively well the observation.
A slight difference appears, however, for the case with cylinders
of diameter $1$ cm at the filling of 0.006. In our results, a
small attenuation peak occurs at about 1.5 kHz, but is absent from
the experiment. This discrepancy may be attributed to a couple of
reasons: the theoretical setting does not match exactly that in
the experiment and perhaps the attenuation peak is too small to be
observable. In Fig.~1 we also observe some peaks located at higher
frequencies as observed in the experiment. These peaks are
sensitive to the arrangement of the transmitter and receiver, and
the number of the cylinders. They are not because the frequencies
are within a stop band.

\input{epsf}
\begin{figure}[hbt]
\begin{center}
\epsfxsize=3in \epsffile{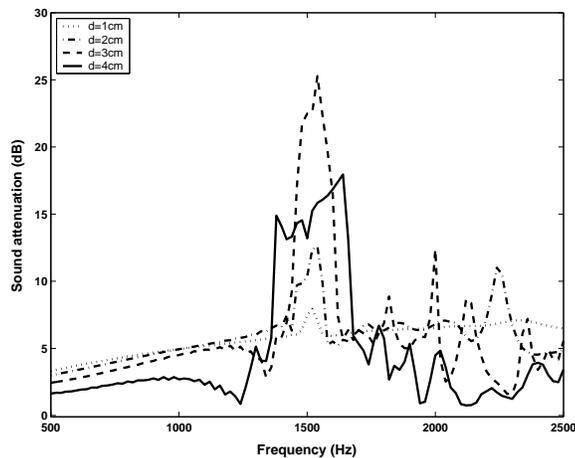} \caption{Theoretical results
for the relative acoustic attenuation along [100] as a function of
frequency for various arrangements of cylinders. The parameters
for the cylinder arrays are taken from the experiment.
}\label{fig1}
\end{center}
\end{figure}

Fig.~2 shows the attenuation (relative) spectra for the case of
cylinders of diameter 3 cm in a square lattice with lattice
constant 11 cm. The corresponding band structure is also depicted.
The attenuation along the [100] and the [110] directions are
represented in solid and dotted lines on the right panel
respectively. Comparing Fig.~2 with the experimental Fig.~2 in
\cite{Sanchez}, we see that the attenuation peak along [100]
coincides almost exactly with the experimental data in the
frequency range between 1.38 and 1.70 kHz. In this particular
case, the attenuation peaks are also roughly in the same order of
magnitude as the observation. Along the [110] direction, the
attenuation due to the deaf bands are also nicely recovered by the
theory. That is, the two bands predicted by the band structure
computation (the second and the third bands in Fig.~2 for the
range from M to L) are actually deaf and wave propagation is
prohibited within these two bands. Similar results have also been
reported for 2D photonic band gap materials\cite{Exp1,pbg}.
Further simulation shows that though the height of the attenuation
peaks may vary as the outer boundary of the cylinder arrays
changes, the overall shape of the attenuation spectra remain
unchanged.

\input{epsf}
\begin{figure}[hbt]
\begin{center}
\epsfxsize=3in \epsffile{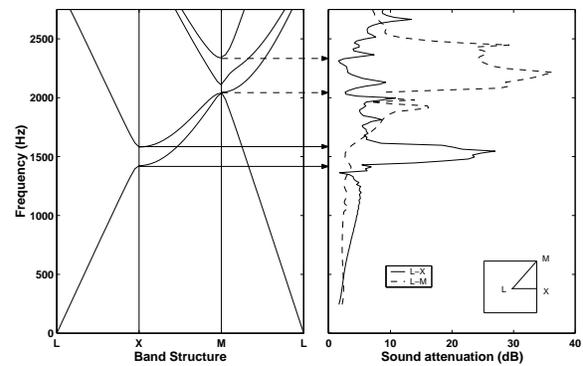} \caption{ Right panel: Relative
acoustic attenuation vs. frequency for a square array of cylinders
with periodicity 11 cm. The radius of the cylinders is 3 cm. Left
panel: The band structures computed by the plane wave expansion
method.}\label{fig2}
\end{center}
\end{figure}

We also notice that the width of the attenuation peak along [110]
is a little wider than the observation. In addition to the
aforementioned reasons, this discrepancy could be due to the fact
that in the present simulation, the cylinders are assumed to be in
the open air, while the experiment was performed in a chamber
which may somewhat still reflect sound. Furthermore, the exact
number and the setting of the cylinders in the experiment are also
not known from the literature. Other contributions to the
discrepancy may result from the different acoustic source and
receiver used in the theory and experiment. In spite of these
limitations, the match between the theory and experiment is quite
encouraging.

The band structure shown in Fig.~2 is obtained by using the usual
plane-wave method\cite{Kush}. Here it is shown to agree nicely
with the band structure obtained by the variation method
calculation. With the parameters in Fig.~2, wave propagation in
different directions is inhibited within different frequency
regimes. No complete band gap is observed. The overlap of
attenuation peaks along different directions, an indication of the
complete band gap, can observed when cylinder filling factor
exceeds certain values\cite{Sanchez}; in this case, the matrix
inversion in the multiple scattering computation becomes costly.

\input{epsf}
\begin{figure}[hbt]
\begin{center}
\epsfxsize=3in \epsffile{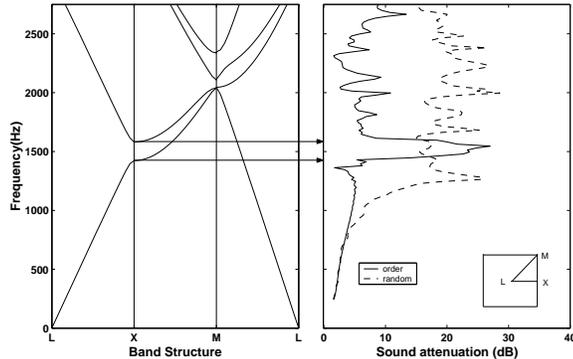} \caption{ The right panel shows
the relative acoustic attenuation vs. frequency. Here the
comparison is made between the results from a square array with
periodicity 11 cm (solid line) and from a complete random array of
cylinders (dotted line). The radius of the cylinders is 3 cm. Left
panel: The band structures computed by the plane wave expansion
method.}\label{fig3}
\end{center}
\end{figure}

Now consider the effect of randomness on the acoustic
transmission. Here we take the case described by Fig.~2 as the
example. While keeping the same cylinder filling and the outer
boundary of the array, we allow the cylinders to be distributed
completely randomly within the area. The relative attenuation is
computed again. The results are shown in Fig.~3, with comparison
to the results for wave propagation along [100] for the
corresponding square latter array. The following features are
evident. At low frequencies, the introduced disorder does not
affect the transmission for the given size of the sample. The
randomness reduces the transmission basically for all frequencies
above a critical frequency of 700 Hz, except for the frequencies
within the lowest stop band along [100] within which though the
transmission is still inhibited, the disorder actually reduces the
attenuation purely due to the stop-band effect; such a feature is
also observed in other acoustic systems\cite{Mara}. As we increase
the sample size, the critical frequency tends to become smaller,
implying larger ranges of inhibition. Note that in order to
compute the transmission accurately at lower frequencies, larger
sample sizes are required and the computation would become costly.
The result of the severe reduction in transmission for a wide
range of frequencies is remarkable and has significant relevance
to the fundamental problem of Anderson localization, a concept
originally introduced to explain the conductor-insulator
transition induced by disorders in electronic systems\cite{Local},
because such a reduction is a precursor to the localization
phenomenon. The results also imply that random arrays of rigid
cylinders are good candidates in filtering audible noise.

In summary, here we have presented a theory for acoustic
transmission through arrays of rigid cylinders in the air. The
theory are applied to the experimental situations of regular
arrays, yielding favorable agreements. The theoretical results
verify the existence of the deaf bands. The results are
subsequently extended to the case of random cylinder arrays. We
found that wave propagation is significantly reduced by randomness
for a wide range of frequencies. This feature makes the rigid
cylinders in the air an ideal system for theoretical and
experimental studies of wave localization.

The help from E. Hoskinson now at UC Berkeley (Physics) is greatly
appreciated. Useful communication with J. Sanchez-Dehesa is also
thanked. The work received support from NSC.

\end{document}